\documentclass{jpsj-suppl}
\usepackage{times}
\usepackage{color}

\title{Ground state of S=1 zigzag spin-orbital chain}

\author{Hiroaki Onishi\thanks{E-mail address: onishi.hiroaki@jaea.go.jp}
}
\inst{Advanced Science Research Center,
Japan Atomic Energy Agency,
Tokai, Ibaraki 319-1195, Japan
}
\abst{%
We investigate ground-state properties of a $t_{\rm 2g}$-orbital Hubbard model
on a zigzag chain relevant for CaV$_{2}$O$_{4}$,
by exploiting numerical techniques
such as Lanczos diagonalization and density-matrix renormalization group.
Assuming a V$^{3+}$ ion,
a local spin $S$=$1$ state is formed by two electrons
in the $t_{\rm 2g}$ orbitals.
That is, the system is a Haldane system
with active $t_{\rm 2g}$-orbital degrees of freedom.
We observe orbital-state transitions,
yielding a distinct spin system under the orbital-ordered background.
We also discuss the orbital structure induced by open edges,
originating in the spatial anisotropy of the $t_{\rm 2g}$ orbitals.
}

\kword{%
Haldane system,
frustration,
orbital degrees of freedom,
CaV$_{2}$O$_{4}$
}

\begin{document}
\maketitle

\section{Introduction}

It is widely recognized that
the interplay of magnetic frustration and quantum fluctuations is
a key ingredient for the emergence of novel magnetism
in quantum spin systems.
One of the most simplest and extensively studied model is
a one-dimensional antiferromagnetic zigzag spin chain
with competing nearest-neighbor and next-nearest-neighbor exchange interactions.
In the case of $S$=$1$,
the ground state of an isotropic Heisenberg chain without frustration
is the so-called Haldane phase,
in which no magnetic long-range order occurs
and a finite energy gap (Haldane gap) opens in the spin excitation.
For the zigzag spin chain,
the ground state turns to be a gapless chiral phase
in a region of strong frustration and easy-plane anisitropy
\cite{Hikihara2000}.
Note that the gapless chiral phase can be regarded as the vestige of
a helical ordered phase
in the classical limit $S$$\rightarrow$$\infty$
\cite{Hikihara2001}.

CaV$_{2}$O$_{4}$,
in which V$^{3+}$ ions with $S$=$1$ form a zigzag structure,
has been studied
as a candidate material for the $S$=$1$ zigzag spin chain.
Since the V-V distances are nearly identical,
nearest-neighbor and next-nearest-neighbor exchanges are
expected to be the same order,
indicating strong frustration.
The early NMR experiments showed a gapless nature,
suggesting a possible realization of the gapless chiral phase
\cite{Fukushima2002}.
However,
recent NMR results gave an evidence for an antiferromagnetic transition
at $T_{\rm N}$=$78$~K
\cite{Zong2008},
consistent with previous neutron diffraction measurements
\cite{Hastings1967}.
Based on susceptibility and neutron diffraction measurements,
it has been suggested that
CaV$_{2}$O$_{4}$ behaves as weakly coupled Haldane chains
at high temperatures above a structural phase transition temperature $T_{\rm S}$=$141$~K,
while it changes to a spin ladder at low temperatures below $T_{\rm S}$
\cite{Pieper2009}.
In an orbital-based senario,
these two spin systems are explained by different orbital configurations,
caused by the structural distortion.
That is, V$^{3+}$ ions have two electrons in $t_{\rm 2g}$ orbitals,
so that $t_{\rm 2g}$-orbital degrees of freedom play a crucial role
in determining the magnetic properties of vanadium systems
\cite{Onishi2004,Khaliullin2005}.
To investigate the physics of frustrated vanadium chains,
the ground state of an effective spin-orbital exchange model has been examined
\cite{Chern2009}.
The effects of
the spin-orbit coupling and the Jahn-Teller distortion have been discussed.

In this paper,
to clarify a key role of active $t_{\rm 2g}$-orbital degrees of freedom
in the $S$=$1$ Haldane system,
we analyze the ground state of a $t_{\rm 2g}$-orbital Hubbard model
relevant for CaV$_{2}$O$_{4}$ by numerical methods.
We observe orbital-state transitions
that cause the change of the spin system,
consistent with the previous work on the spin-orbital exchange model
\cite{Chern2009}.
We also discuss the orbital structure
in a lattice with open edges,
appearing due to the spatial anisotropy of $t_{\rm 2g}$ orbitals.

\section{Model and method}

Let us consider the $t_{\rm 2g}$ orbitals
on each site of a zigzag chain with $N$ sites.
We assume that the site represents a $V^{3+}$ ion
and the zigzag structure originates in edge-sharing VO$_{6}$ octahedra
in CaV$_{2}$O$_{4}$.
The Hamiltonian of a $t_{\rm 2g}$-orbital Hubbard model is given by
\begin{eqnarray}
 H
 &=&
 -\sum_{{\bf i},{\bf a},\gamma,\gamma',\sigma}
 t_{\gamma\gamma'}^{{\bf a}}
 d_{{\bf i}\gamma\sigma}^{\dag} d_{{\bf i}+{\bf a}\gamma'\sigma}
 +U \sum_{{\bf i},\gamma}
 \rho_{{\bf i}\gamma\uparrow} \rho_{{\bf i}\gamma\downarrow}
 +\frac{U'}{2} \sum_{{\bf i},\sigma,\sigma',\gamma\neq\gamma'} 
 \rho_{{\bf i}\gamma\sigma} \rho_{{\bf i}\gamma'\sigma'}
\nonumber\\
&&
 +\frac{J}{2} \sum_{{\bf i},\sigma,\sigma',\gamma\neq\gamma'} 
 d_{{\bf i}\gamma\sigma}^{\dag} d_{{\bf i}\gamma'\sigma'}^{\dag}
 d_{{\bf i}\gamma\sigma'} d_{{\bf i}\gamma'\sigma}
 +\frac{J'}{2} \sum_{{\bf i},\sigma\neq\sigma',\gamma\neq\gamma'} 
 d_{{\bf i}\gamma\sigma}^{\dag} d_{{\bf i}\gamma\sigma'}^{\dag}
 d_{{\bf i}\gamma'\sigma'} d_{{\bf i}\gamma'\sigma}
\nonumber\\
&&
 -\frac{\Delta}{3} \sum_{{\bf i}}
 (2\rho_{{\bf i}xy}-\rho_{{\bf i}yz}-\rho_{{\bf i}zx})
 -\frac{\Delta'}{2} \sum_{{\bf i}}
 (\rho_{{\bf i}yz}-\rho_{{\bf i}zx}),
\label{eq1}
\end{eqnarray}
where $d_{{\bf i}\gamma\sigma}$ is an annihilation operator
for an electron
with spin $\sigma$ (=$\uparrow$, $\downarrow$)
in orbital $\gamma$ (=$xy,yz,zx$)
at site ${\bf i}$,
$\rho_{{\bf i}\gamma\sigma}$=$d_{{\bf i}\gamma\sigma}^{\dag}d_{{\bf i}\gamma\sigma}$,
and $\rho_{{\bf i}\gamma}$=$\sum_{\sigma}\rho_{{\bf i}\gamma\sigma}$.
For the hopping term,
$t_{\gamma\gamma'}^{{\bf a}}$ is the hopping amplitude
between $\gamma$ and $\gamma'$ orbitals in adjacent sites
along ${\bf a}$ (=${\bf u},{\bf v},{\bf w}$, see Fig.~1(a)).
Since the V-O-V bond angle is nealy $90^{\circ}$ in CaV$_{2}$O$_{4}$,
we only consider the direct $t_{\rm 2g}$-orbital hopping through the $\sigma$ bond
\cite{Chern2009}.
The hopping amplitude is evaluated from the overlap integral of
the $t_{\rm 2g}$-orbital wavefunctions
\cite{Slater1954},
given by
$t_{xy,xy}^{{\bf u}}$=$t_{yz,yz}^{{\bf v}}$=$t_{zx,zx}^{{\bf w}}$=$t$
and zero for other combinations of orbitals and directions.
Hereafter, we set $t$=$1$ as the energy unit.
Regarding the local interactions,
$U$, $U'$, $J$, and $J'$ are
the intra-orbital Coulomb repulsion,
the inter-orbital Coulomb repulsion,
the inter-orbital exchange interaction (Hund's rule coupling),
and the pair-hopping interaction,
respectively.
We assume the relations $U$=$U'$+$J$+$J'$ and $J$=$J'$.
For the crystalline electric field effects,
$\Delta$ represents the tetragonal distortion of VO$_{6}$ octahedra,
while
$\Delta'$ denotes the orthorhombic distortion,
leading to the level splitting of the $t_{\rm 2g}$ level,
as shown in Fig.~1(b).
Note that $\Delta$ is supposed to be positive for CaV$_{2}$O$_{4}$,
while $\Delta'$ changes due to the structural phase transition
from zero in the high-temperature phase
to small positive in the low-temperature phase.
In the present study,
we fix $U'$=$10$ and $J$=$1$,
and investigate the dependence on $\Delta$ and $\Delta'$.

\begin{figure}[t]
\begin{center}
\includegraphics[scale=0.8]{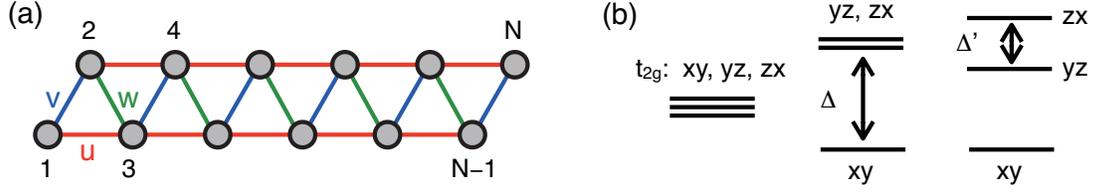}
\end{center}
\caption{(Color online)
(a) The zigzag chain structure of CaV$_{2}$O$_{4}$.
The sites are numbered along the zigzag path.
The hopping amplitude $t_{\gamma\gamma'}^{{\bf a}}$ has a finite matrix element $t$
between $xy$ orbitals in adjacent sites along ${\bf u}$,
$yz$ orbitals along ${\bf v}$,
and
$zx$ orbitals along ${\bf w}$.
(b) The level splitting due to the crystalline electric field.
}
\label{fig1}
\end{figure}

We analyze the ground state of the model (\ref{eq1})
by numerical techniques.
We mainly use the Lanczos diagonalization method.
Note that because of the three orbitals in each site,
the matrix dimension of the Hamiltonian becomes huge
as the system size increases.
Indeed,
considering the subspace of $S_{\rm tot}^{z}$=$0$,
where $S_{\rm tot}^{z}$ is the $z$-component of the total spin,
the matrix dimension is $245 \, 025$ for $N$=$4$,
while it grows to $344 \, 622 \, 096$ for $N$=$6$,
and $540 \, 917 \, 591 \, 841$ for $N$=$8$.
In this paper, we deal with a small periodic chain with $N$=$4$
to obtain results with reasonably short CPU time.
For the analysis of a large system with $N$=$12$,
we also exploit the density-matrix renormalization group (DMRG) method
\cite{White1992}.
We adopt the finite-system algorithm
with open boundary conditions.
In the present calculations,
the number of kept states is up to $m$=$200$,
and the truncation error is around $10^{-4}$.

\section{Numerical results}

\begin{figure}[t]
\begin{center}
\includegraphics[scale=0.8]{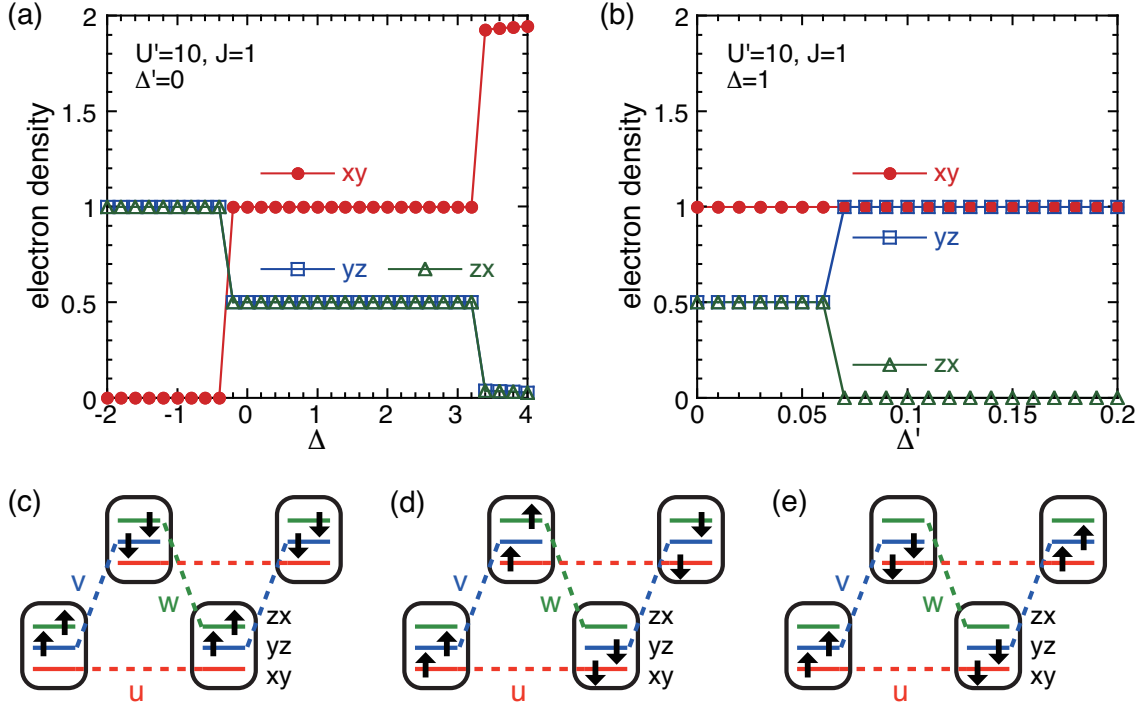}
\end{center}
\caption{(Color online)
Lanczos results for the $4$-site periodic chain.
(a) The $\Delta$ dependence of the electron density in each orbital
at $U'$=$10$, $J$=$1$, and $\Delta'$=$0$.
(b) The $\Delta'$ dependence of the electron density in each orbital
at $U'$=$10$, $J$=$1$, and $\Delta$=$1$.
The electron configuration in different spin-orbital states is depicted in (c)-(e):
(c) $\Delta$$\lesssim$$0$ and $\Delta'$$\simeq$$0$.
(d) $0$$\lesssim$$\Delta$$\lesssim$$3J$ and $\Delta'$$\simeq$$0$.
(e) $0$$\lesssim$$\Delta$$\lesssim$$3J$ and $\Delta'$$\gtrsim$$0.06$.
}
\label{fig2}
\end{figure}

To gain an insight into the orbital state,
we measure the electron density in each orbital,
\begin{equation}
 n_{\gamma} = \frac{1}{N}\sum_{{\bf i}} \langle \rho_{{\bf i}\gamma} \rangle,
\end{equation}
where $\langle \cdots \rangle$ denotes the expectation value
in the ground state.
In Fig.~2(a), we show the $\Delta$ dependence of $n_{\gamma}$
at $U'$=$10$, $J$=$1$, and $\Delta'$=$0$.
When $\Delta$ is negative,
the lower $yz$ and $zx$ orbitals are singly occupied,
indicating a ferro-orbital (FO) state,
while the upper $xy$ orbital is vacant.
In Fig.~2(c), we depict the spin-orbital configuration.
Note that the Hund's rule coupling stabilizes a local spin $S$=$1$ state,
described by parallel spins in the $yz$ and $zx$ orbitals in each site.
Here, we discuss the spin exchange interaction
through the virtual electron hopping process between adjacent sites
with an underlying orbital configuration.
Electrons in the $yz$ orbitals can hop along the ${\bf v}$ direction.
Considering the second-order process of the electron hopping,
an antiferromagnetic (AFM) interaction should occur.
In the same manner,
electrons in the $zx$ orbitals can hop along the ${\bf w}$ direction,
and the second-order process of the electron hopping yields an AFM interaction.
In contrast, there is no electron in the $xy$ orbitals,
so that no spin exchange interaction arises along the ${\bf u}$ direction.
Consequently,
the system is regarded as a spin $S$=$1$ AFM chain
along the zigzag path.

With increasing $\Delta$,
electrons tend to occupy the lower $xy$ orbitals
when $\Delta$ becomes positive.
In fact,
we find an orbital-state transition at around $\Delta$$\simeq$$0$,
as shown in Fig.~2(a).
The lower $xy$ orbital is singly occupied,
indicating an $(xy)$-type FO configuration.
On the other hand, the upper $yz$ and $zx$ orbitals are equally occupied by one electron.
In such a case, it is useful to introduce a pseudospin $T$=$\frac{1}{2}$
to describe the orbital state, such that
$T_{{\bf i}}^{z}$=$\frac{1}{2}$ for the $yz$ orbital
and $T_{{\bf i}}^{z}$=$-$$\frac{1}{2}$ for the $zx$ orbital,
where $T_{{\bf i}}^{z}$=$\frac{1}{2}\sum_{\sigma}(d_{{\bf i}yz\sigma}^{\dag}d_{{\bf i}yz\sigma}$$-$$d_{{\bf i}zx\sigma}^{\dag}d_{{\bf i}zx\sigma})$.
We observe that
$\langle T_{{\bf i}}^{z} \rangle$ is zero at every site,
since the $yz$ 
and $zx$ orbitals are equally occupied.
To clarify the orbital state,
we measure the orbital correlation function
$\langle T_{{\bf i}}^{z} T_{{\bf i}}^{z} \rangle$,
and it turns out that a $(yz/zx)$-type antiferro-orbital (AFO) correlation is robust.
That is, electrons occupy the $yz$ or $zx$ orbital alternately along the zigzag path.
In Fig.~2(d), we show the spin-orbital configuration.
Let us here discuss the spin exchange interaction.
Since we have the $(xy)$-type FO configuration,
the second-order hopping process
between the adjacent $xy$ orbitals along the ${\bf u}$ direction
leads to an AFM interaction.
On the other hand,
due to the $(yz/zx)$-type AFO configuration along the zigzag path,
the second-order hopping process
between the adjacent $yz$ $(zx)$ orbitals along the ${\bf v}$ $({\bf w})$ direction
leads to a ferromagnetic (FM) interaction.
Therefore, the system is considered as a spin $S$=$1$ zigzag chain
with nearest-neighbor FM and next-nearest-neighbor AFM interactions,
corresponding to the high-temperature phase
of CaV$_{2}$O$_{4}$
\cite{Pieper2009,Chern2009}.

Note that the double occupancy is prohibited
due to the intra-orbital Coulomb repulsion,
when $\Delta$ is moderately small.
However, if we further increase $\Delta$,
two electrons are forced to occupy the lower $xy$ orbital.
As shown in Fig.~2(a),
we find an orbital-state transition at around $\Delta$$\simeq$$3J$.
The transition point is roughly estimated by comparing the diagonal part of the local energy.
The local energy is given by $U'$$-$$J$$+$$\Delta$
below the transition point,
in which the $xy$ orbital is singly occupied and the $yz$
and $zx$ orbitals are equally occupied.
Above the transition point,
the local energy is $U$, since the $xy$ orbital is doubly occupied.
Comparing these energies with noting the relations $U$=$U'$+$J$+$J'$ and $J$=$J'$,
the transition point is estimated to be $3J$,
consistent with the numerical results.

\begin{figure}[t]
\begin{center}
\includegraphics[scale=0.8]{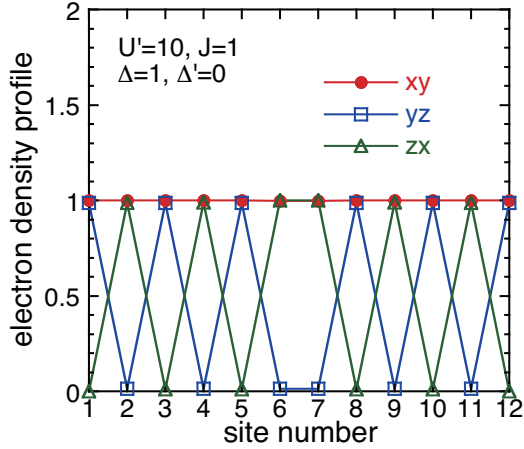}
\end{center}
\caption{(Color online)
DMRG results for the $12$-site open chain
with the lattice configuration in Fig.~1(a).
The electron density profile
at $U'$=$10$, $J$=$1$, and $\Delta'$=$0$.
}
\label{fig3}
\end{figure}

Here, let us consider the effects of $\Delta'$.
In Fig.~2(b), we show the $\Delta'$ dependence of $n_{\gamma}$
at $U'$=$10$, $J$=$1$, and $\Delta$=$1$.
We find an orbital-state transition at $\Delta'$$\simeq$$0.06$,
above which the $xy$ and $yz$ orbitals are singly occupied,
indicating a FO state.
Regarding the spin exchange interaction,
we have an AFM interaction
through the second-order hopping process
between the adjacent $xy$ $(yz)$ orbitals along the ${\bf u}$ $({\bf v})$ direction,
while there is no spin exchange interaction along the ${\bf w}$ direction.
That is, the system is a spin $S$=$1$ AFM ladder,
corresponding to the low-temperature phase
of CaV$_{2}$O$_{4}$
\cite{Pieper2009,Chern2009}.

Now we move to the DMRG results.
Note that we adopt open boundary conditions in the DMRG calculations,
since, in general,
we obtain precise results with open boundary conditions
compared with periodic boundary conditions
within the same computational cost.
However, we should give careful consideration to the boundary effects.
In the present case,
the hopping amplitude depends on the direction and the occupied orbitals
due to the spatial anisotropy of the $t_{\rm 2g}$ orbitals,
so that open edges introduce a special type of bond configuration.
In fact, we can clearly see the boundary effects
in the electron density profile, defined by
\begin{equation}
 n_{{\bf i}\gamma} = \langle \rho_{{\bf i}\gamma} \rangle.
\end{equation}
In Fig.~3, we show $n_{{\bf i}\gamma}$
at $U'$=$10$, $J$=$1$, $\Delta$=$1$, and $\Delta'$=$0$.
We find that the $xy$ orbital is singly occupied at every site,
while the $yz$ and $zx$ orbitals are alternately occupied along the zigzag path,
and a kink of the alternating pattern appears at the center of the system.
This is because the edge site is connected to the nearest-neighbor site
through the bond along the ${\bf v}$ direction,
where the electron hopping is effective for the $yz$ orbital
and the $zx$ orbital is localized.
To gain the kinetic energy,
the $yz$ orbital is preferably occupied rather than the $zx$ orbital at the edge sites.
On the other hand,
if the lattice has a bond along the ${\bf w}$ direction at an open edge,
the $zx$ orbital is itinerant and favorably occupied at the edge site.
Thus,
we observe an edge-induced orbital structure
due to the spatial anisotropy of the $t_{\rm 2g}$ orbitals.
In sharp contrast,
as we have discussed above,
Lanczos results for the periodic chain show that
$n_{{\bf i}xy}$$\simeq$$1$ and $n_{{\bf i}yz}$=$n_{{\bf i}zx}$$\simeq$$0.5$,
and we can detect an alternating orbital configuration
only in terms of the orbital correlation.
To avoid the boundary effects,
we are performing DMRG calculations with periodic boundary condistions,
which will be reported elsewhere.

\section{Summary}

We have studied ground-state properties of
the $t_{\rm 2g}$-orbital Hubbard model on the zigzag chain
relevant for vanadates by numerical methods.
We have observed orbital-state transitions
that lead to the change of the spin system.
According to the orbital-ordered background,
the system can be regarded as an antiferromagnetic spin chain,
a zigzag spin chain with ferromagnetic nearest-neighbor
and antiferromagnetic next-nearest-neighbor exchanges,
or an antiferromagnetic spin ladder.
We have also shown the edge-induced orbital structure,
caused by the spatial anisitropy of the $t_{\rm 2g}$ orbitals.

\section*{Acknowledgement}

The author is grateful to N. Todoroki for discussions.
He also thanks S. Maekawa, M. Mori and T. Sugimoto for discussions and comments.
Part of calculations were done on the supercomputer at the Japan Atomic Energy Agency.





\begin{thebibliography}{9}

\bibitem{Hikihara2000}
T. Hikihara, M. Kaburagi, H. Kawamura, and T. Tonegawa:
J. Phys. Soc. Jpn. {\bf 69} (2000) 259.

\bibitem{Hikihara2001}
T. Hikihara, M. Kaburagi, and H. Kawamura:
Phys. Rev. B {\bf 63} (2001) 174430.

\bibitem{Fukushima2002}
H. Fukushima, H. Kikuchi, M. Chiba, Y. Fujii, Y. Yamamoto,
and H. Hori:
Prog. Theor. Phys. Suppl. {\bf 145} (2002) 72.

\bibitem{Zong2008}
X. Zong, B. J. Suh, A. Niazi, J. Q. Yan, D. L. Schlagel,
T. A. Lograsso, and D. C. Johnston:
Phys. Rev. B {\bf 77} (2008) 014412.

\bibitem{Hastings1967}
J. M. Hastings, L. M. Corliss, W. Kunnmann, and S. La Placa:
J. Phys. Chem. Solids {\bf 28} (1967) 1089.

\bibitem{Pieper2009}
O. Pieper, B. Lake, A. Daoud-Aladine, M. Reehuis, K. Proke\v{s},
B. Klemke, K. Kiefer, J. Q. Yan, A. Niazi, D. C. Johnston,
and A. Honecker:
Phys. Rev. B {\bf 79} (2009) 180409(R).

\bibitem{Onishi2004}
H. Onishi and T. Hotta:
Phys. Rev. B {\bf 70} (2004) 100402(R).

\bibitem{Khaliullin2005}
G. Khaliullin:
Prog. Theor. Phys. Suppl. {\bf 160} (2005) 155.

\bibitem{Chern2009}
G.-W. Chern and N. Perkins:
Phys. Rev. B {\bf 80} (2009) 220405(R).

\bibitem{Slater1954}
J. C. Slater and G. F. Koster:
Phys. Rev. {\bf 94} (1954) 1498.

\bibitem{White1992}
S. R. White:
Phys. Rev. Lett. {\bf 93} (1992) 2863.

\end{thebibliography}
\end{document}